\documentclass[reprint, showpacs, superscriptaddress, aps, prl, twocolumn, 10pt]{revtex4-1}

\usepackage{graphicx} 
\usepackage{dcolumn}
\usepackage{bm}
\usepackage{amssymb}
\usepackage{amsmath}
\usepackage{wasysym}
\usepackage{color}
\usepackage{hyperref}
\usepackage{dcolumn}
\usepackage{float}
\usepackage{flushend}
\hypersetup{
colorlinks,
citecolor=blue,
filecolor=blue,
linkcolor=blue,
urlcolor=blue}

\begin{document}
\title{Energy spectrum of Buoyancy-driven Flows}
\author{Abhishek Kumar}
\affiliation {Department of Physics, Indian Institute of Technology - Kanpur, India 208016}
\author{Anando G. Chatterjee}
\affiliation {Department of Physics, Indian Institute of Technology - Kanpur, India 208016}
\author{Mahendra K. Verma}
\email{mkv@iitk.ac.in}
\affiliation {Department of Physics, Indian Institute of Technology - Kanpur, India 208016}

\begin{abstract}
Using high-resolution direct numerical simulation and arguments based on the kinetic energy flux $\Pi_u$, we demonstrate  that for stably stratified flows, the kinetic energy spectrum $E_u(k) \sim k^{-11/5}$,  the entropy spectrum $E_\theta(k) \sim k^{-7/5}$, and  $\Pi_u(k) \sim k^{-4/5}$ (Bolgiano-Obukhov scaling). This scaling is due to the depletion of kinetic energy because of buoyancy.   For weaker buoyancy in stratified flows, $E_u(k)$ follows Kolmgorov's spectrum with a constant energy flux.  We also argue that for Rayleigh B\'{e}nard convection, the Bolgiano-Obukhov scaling will not hold for the bulk flow due to the positive energy supply by buoyancy and non-decreasing $\Pi_u(k)$.

\end{abstract}

\maketitle
Buoyancy or  density gradient drives flows in the atmosphere and interiors of planets and stars, as well as in electronic devices and industrial applications like heat exchangers, boilers, etc.  Accordingly, scientists (including geo-, astro-, atmospheric- and solar physicists) and engineers have been working on understanding buoyancy driven flows for more than a century.  An important unsolved problem in this field is how to quantify the spectra and fluxes of kinetic energy (KE) and entropy ($u^2/2$ and $\theta^2/2$ respectively, where $\mathbf u$ and $\theta$ are the velocity and temperature fluctuations) of buoyancy driven flows~\cite{Siggia:ARFM1994,Lohse:ARFM2010}. In this letter, we will study these quantities and respective nonlinear fluxes using direct numerical simulations, and show that the spectrum differs from  Kolmogorov's theory when buoyancy is strong. 

Flows driven by buoyancy can be classified in two categories: (a) convective flows in which hotter and lighter fluid at the bottom rises, while  colder and heavier fluid at the top comes down.  These flows are unstable;  (b) Stably stratified flows in which lighter fluid rests above  heavier fluid.    Stably stratified flows are  stable, hence their fluctuations vanish over time.  Therefore, they need to be driven by an external force to obtain a steady turbulent state.  Even though both types of flows are driven by  density gradients, the properties of such flows are quite different, which we decipher using quantitative analysis of energy flux and energy  supply rate by buoyancy.

\begin{figure}[htbp]
\begin{center}
\includegraphics[scale = 1.0]{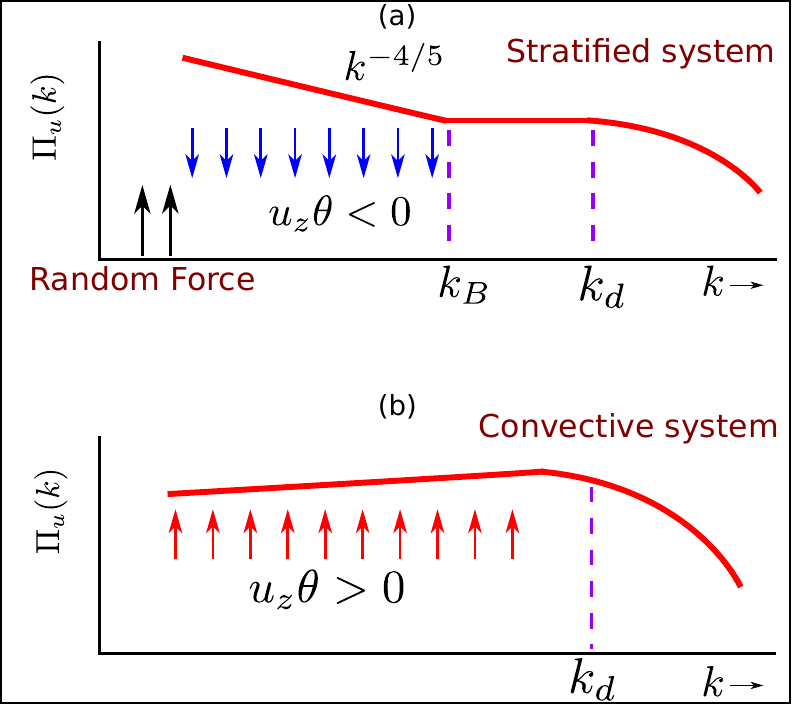}
\end{center}
\caption{Schematic diagrams of energy flux $\Pi_u(k)$: (a) In a stably stratified flow,  $\Pi_u(k)$ decreases with $k$  due to a negative energy supply rate $\Re \langle u_z(k) \theta^*(k)\rangle$.  (b) In a thermally driven flow (e.g., Rayleigh B\'{e}nard convection), $\Pi_u(k)$  is a non-decreasing function of $k$ due to positive $\Re \langle u_z(k) \theta^*(k) \rangle$.  }
\label{fig:sch_flux}
\end{figure}

For stably stratified flows, Bolgiano~\cite{Bolgiano:JGR1959} and Obukhov~\cite{Obukhov:DANS1959} first proposed a phenomenology, according to which the KE flux $\Pi_u$ of a stably stratified flow is depleted at different length scales due to conversion of kinetic energy to ``potential energy'' via buoyancy ($u_z \theta$).  As a result, $\Pi_u(k)$ decreases with wavenumber (see Fig.~\ref{fig:sch_flux}{\color{blue}(a)}), and the energy spectrum is steeper than that prediced by Kolmogorov theory $(E(k) \sim k^{-5/3}$, where $k$ is the wavenumber). According to the phenomenology proposed by Bolgiano and Obukbhov (referred to as {\em BO}), for $k<k_B$ ($k_B = $ Bolgiano wavenumber~\cite{Bolgiano:JGR1959}), the KE spectrum $E_u(k)$, entropy spectrum $E_\theta(k)$,  $\Pi_u$, and entropy flux $\Pi_\theta$ are:
\begin{eqnarray}
E_u(k) & =  & c_1 (\alpha^2 g ^2 \epsilon_\theta)^{2/5}k^{-11/5}, \label{eq:Eu} \\
E_\theta(k) & =  & c_2 (\alpha g)^{-2/5}\epsilon_\theta^{4/5} k^{-7/5}, \label{eq:Etheta} \\
\Pi_\theta(k) & = &  \epsilon_\theta = \mathrm{constant}, \label{eq:pi_theta} \\
\Pi_u(k) & = & c_3 (\alpha^2 g^2 \epsilon_\theta)^{3/5} k^{-4/5},  \label{eq:pi}
\end{eqnarray}
where $\alpha$, $g$, and $\epsilon_{\theta}$ are the thermal expansion coefficient, acceleration due to gravity, and the entropy dissipation rate respectively, and $c_i$'s are constants.    

According to the {\em BO} theory, the decrease in $\Pi_u(k)$ occurs due to a  negative energy supply rate $F(k) = \Re \langle u_z(k) \theta^*(k)\rangle$, where $\Re$ stands for the real part of the argument.    For the wavenumbers in the range $k_B < k < k_d$,  $\Pi_u \approx \epsilon_u$, and $E_u(k), E_\theta(k) \sim k^{-5/3}$   (see Fig.~\ref{fig:sch_flux}{\color{blue}(a)}).  Here $\epsilon_u$ is the KE dissipation rate, and $k_d$ is the wavenumber after which dissipation starts.  In this letter we numerically compute $F(k)$ and $\Pi_u(k)$ for stratified turbulence, and show a consistency with the {\em BO} scaling.   We remark that many researchers describe the stably stratified flows in terms of density fluctuation $\rho'$, which leads to an equivalent description since $\theta^2/2$ is proportional to $\rho'^2/2$ (usually referred to as ``potential energy''~\cite{Davidson:book_2}).

Procaccia and Zeitak~\cite{Procaccia:PRL1989}, L'vov~\cite{Lvov:PRL1991},  L'vov and Falkovich~\cite{Lvov:PD1992}, and Rubinstein~\cite{Rubinstein:NASA} proposed that a similar scaling is applicable to Rayleigh-B\'{e}nard convection (RBC).   Their arguments hinges on an assumption that $F(k) = \Re \langle u_z(k) \theta^*(k)\rangle < 0$ even for RBC. Note that for the inertial range regime, under a steady state, the variation of energy flux is given by
\begin{equation}
\frac{d}{dk} \Pi(k) =  F(k) - D(k),
\label{eq:Pi_k_RBC}
\end{equation}
where   $D(k)$ is the viscous dissipation~\cite{Lvov:PRL1991,Lvov:PD1992,Lesieur:book,Verma:EPL2012,suppl}.  In this letter, we demonstrate using numerical simulations that $F(k) > 0$  for RBC [see Fig.~\ref{fig:sch_flux}(b)], unlike stably stratified flows where $F(k) < 0$.  Consequently $\Pi(k)$ would increase with $k$, and $E(k)$ would be either Kolmogorov-like ($E(k) \sim k^{-5/3}$) or shallower ($E(k) \sim k^{-a}$ with $a < 5/3$).  These observations of non-decreasing $\Pi(k)$ contradict the earlier predictions on RBC~\cite{Procaccia:PRL1989,Lvov:PRL1991,Lvov:PD1992}, but they are in agreement with the numerical results of \v{S}kandera et al.~\cite{Skandera:HPCISEG2SBH2009}.   

In the past there have been several attempts to verify {\em BO} scaling for stably stratified flows.   Kimura and Herring~\cite{Kimura:JFM1996}  reported {\em BO} scaling for a narrow band of wavenumbers in their $128^3$ decaying spectral simulation; the Richardson number of their simulations was greater than unity.    Later, Kimura et al.~\cite{Kimura:JFM2012},  Lindborg~\cite{Lindborg:GRL2005,Lindborg:JFM2006}, and Vallgren et al.~\cite{Vallgren:PRL2011} focussed on anisotropic energy spectrum, and attempted to explain $k^{-3}$ KE spectrum observed for the synoptic scale of terrestrial atmosphere.

For  Rayleigh B\'{e}nard convection (RBC), which is a  class of thermally-driven convection, the numerical and experimental results are largely inconclusive.   Based on simulations with periodic boundary conditions, Borue and Orszag~\cite{Borue:JSC1997} and \v{S}kandera et al.~\cite{Skandera:HPCISEG2SBH2009} reported {\em KO} scaling. \v{S}kandera et al.~\cite{Skandera:HPCISEG2SBH2009} reported a constant KE flux, somewhat consistent with the aforementioned argument [Eq.~(\ref{eq:Pi_k_RBC})]. Calzavarini et al.~\cite{Calzavarini:PRE2002} reported {\em BO} scaling in the boundary layer, and {\em KO} scaling in the bulk.   Mishra and Verma~\cite{Mishra:PRE2010} reported {\em KO} scaling for zero- and low Prandtl number flows since $F(k) = \Re \langle u_z(k) \theta^*(k)\rangle$ is active only at low wavenumbers for such flows.    Camussi and Verzicco~\cite{Camussi:EJMF2004,Verzicco:JFM2003} however reported {\em BO} scaling.  The experimental results~\cite{Niemela:NATURE2000} are  more divergent  with some reporting {\em BO} scaling, and some reporting {\em KO} scaling.

In this letter, we focus on Boussinesq stably stratified flows, whose  equations  in a non-dimensionalised form  are
\begin{eqnarray}
\frac{\partial \bf u}{\partial t} + (\bf u \cdot \nabla) \bf u & = & -\nabla \sigma + \mathrm{Gr}\mathrm{Pr}^2 \theta \hat{z} + \mathrm{Pr} \nabla^2 \bf u + \bf f^u, \label{eq:u_ndim} \\
\frac{\partial \theta}{\partial t} + (\bf u \cdot \nabla) \theta & = & -  u_z + \nabla^2 \theta, \label{eq:th_ndim} \\
\nabla \cdot \bf u & = & 0 \label{eq:inc_ndim}, 
\end{eqnarray}
where  $\mathrm{Pr = \nu / \kappa}$ is the Prandtl number, and  $\mathrm{Gr} = \alpha g \frac{dT}{dz}d^4 / \nu^2$ is the Grashof number, which is a ratio of the buoyancy and dissipation terms.   Another important non-dimensional number is Richardson number $\mathrm{Ri} = \alpha g \frac{dT}{dz}d^2 / u_{\rm rms}^2$, which is a ratio of the buoyancy and the nonlinear term $(\bf u \cdot \nabla) \bf u$.  We demonstrate that {\em BO} scaling is observed when $\mathrm{Ri} = O(1)$, but Kolmogorov scaling $E(k) \sim k^{-5/3}$ [referred to as Kolmogorov-Obukhov ({\em KO}) scaling] is observed when $\mathrm{Ri} \ll 1$, or when buoyancy is negligible. 

To test whether {\em BO} scaling is valid or not for the stably stratified flows, we perform direct numerical simulation of  Eqs.~(\ref{eq:u_ndim}-\ref{eq:inc_ndim}) using pseudospectral code Tarang~\cite{Verma:Pramana2013} in a three-dimensional  box of size $(2\pi)^3$.  We employ  periodic boundary condition on all sides~\cite{Kimura:JFM2012}.  We use fourth-order Runge-Kutta (RK4) method  for time stepping,  Courant-Freidricks-Lewey (CFL) condition for computing time step $\Delta t$, and $3/2$ rule  for dealiasing.   To obtain a steady turbulent flow, we apply a random force to the flow in the band $2 \le k \le 4$ using the scheme of Kimura and Herring~\cite{Kimura:JFM2012}.  

 We perform  large-resolution simulations for  $\mathrm Pr =1$ (close to that of air) and  Richardson numbers  $\mathrm{Ri}=4 \times 10^{-7}, 0.01$, and  $0.5$.  Simulations for $\mathrm{Ri}= 0.01$ have been performed on $1024^3$ grid, while that for $4 \times 10^{-7}$ and $0.5$ have been performed on $512^3$ grid.   The parameters of our runs are listed in Table~\ref{table:simulation_details}. All our simulations are fully resolved since $k_{\mathrm max} \eta >1$, where $k_{\mathrm max}$ is the maximum wavenumber of the run, and $\eta$ is the Kolmogorov length scale. 

\begin{table*}
\begin{center}
\caption{Parameters of our numerical simulations: Richardson number $\mathrm{Ri}$, Grashof number $\mathrm{Gr}$, Prandtl number $\mathrm{Pr}$, Grid size, kinetic energy dissipation rate $\epsilon_u$, thermal dissipation rate $\epsilon_{\theta}$, Reynolds number $\mathrm{Re}$, Reynolds number $\mathrm{Re_{\lambda}}$ based on Taylor micro-scale, $k_{max}\eta$ where $\eta$ is the Kolmogorov length, Bolgiano wavenumber $k_B$, and averaged $\Delta t$.}
\begin{tabular}{p{1.6cm} p{1.6cm} p{0.8cm} p{1.2cm} p{1.8cm} p{1.2cm} p{1.2cm} p{1.2cm} p{1.2cm} p{1.2cm} p{1.8cm}}
\hline \hline \\[0.3 pt]
$\mathrm{Ri}$ & $\mathrm{Gr}$ & $\mathrm{Pr}$  & Grid & $\epsilon_u$ & $\epsilon_{\theta}$ &$\mathrm{Re}$ &$\mathrm{Re_\lambda}$& $k_{max} \eta$ & $k_B$ & $\Delta t$\\[2 mm]
\hline \\[0.5 pt]
$0.5$ & $1 \times 10^5$ & $1$ & $512^3$ & $1.4 \times 10^7$ & $60.7$ & $467$ &$220$ &$4.2$ & $6.0$ & $2.5 \times 10^{-5}$\\
$0.01$ & $5 \times 10^3$ & $1$  & $1024^3$ & $4.0 \times 10^7$ & $150$ & $649$ &$260$ &$6.4$ & $8.5$ & $3.5 \times 10^{-6}$\\
$4 \times 10^{-7}$ & $0.1$ & $1$  & $512^3$ & $2.1 \times 10^7$ & $141$ & $510$ &$220$ &$3.8$ & $< 1$ & $2.6 \times 10^{-6}$\\
\hline \hline
\end{tabular}
\label{table:simulation_details}
\end{center}
\end{table*}

\begin{figure}[htbp]
\begin{center}
\includegraphics[scale = 1.0]{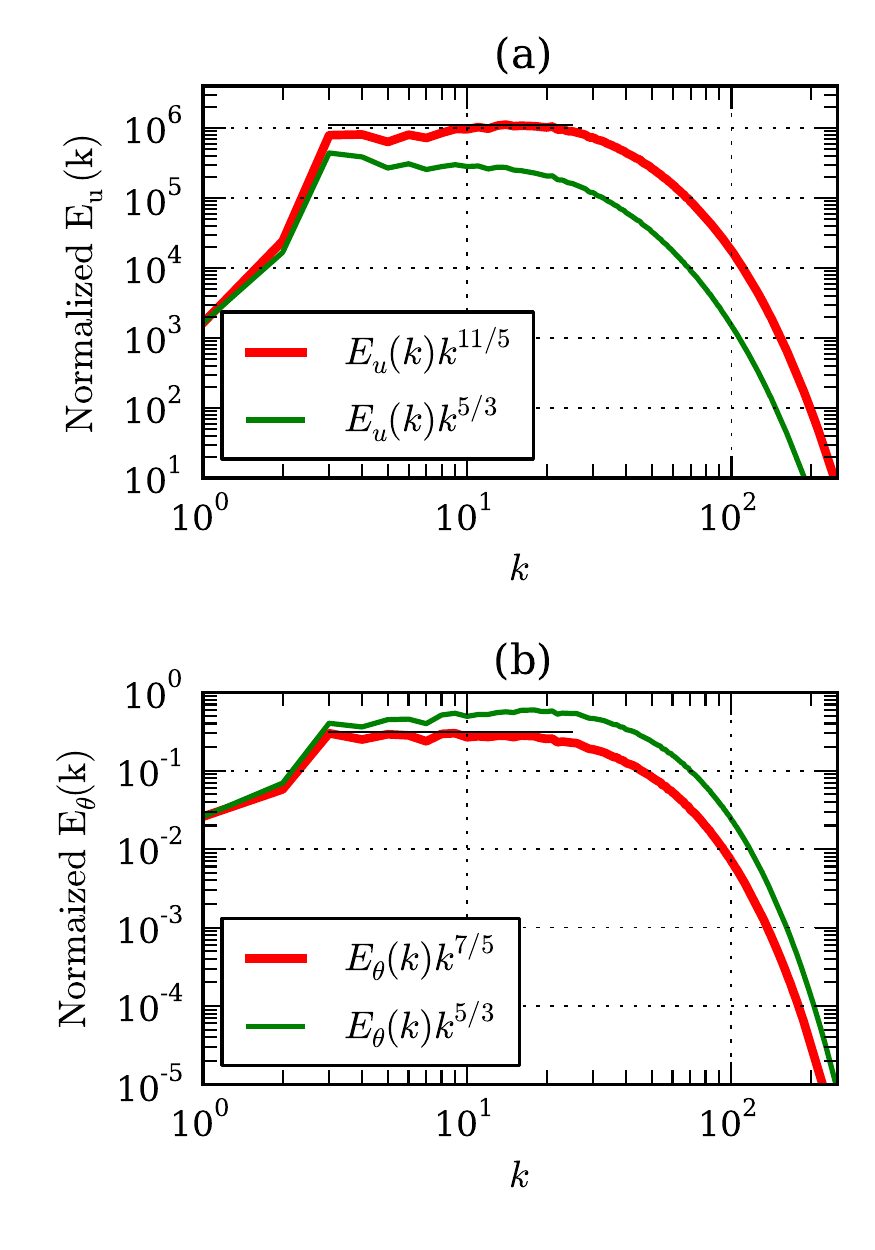}
\end{center}
\caption{For $\mathrm{Pr}=1$ and $\mathrm{Ri} = 0.01$, plots of (a) normalized KE spectra and (b) normalized entropy spectra for Bolgiano-Obukhov ({\em BO}) and Kolmogorov-Obukhov ({\em KO}) scaling.  {\em BO} scaling fits with the data better than {\em KO} scaling.}
\label{fig:spectra_0_01}
\end{figure}

We compute the KE and entropy spectra and fluxes for the steady-state data of $\mathrm{Pr}=1$ and $\mathrm{Ri}=0.01$ run.   In  Fig.~\ref{fig:spectra_0_01}{\color{blue}(a)} we plot the normalized KE spectra, $E_u(k)k^{11/5}$ ({\em BO} scaling) and $E_u(k)k^{5/3}$ ({\em KO} scaling).  In Fig.~\ref{fig:spectra_0_01}{\color{blue}(b)} we plot the normalized entropy spectra, $E_\theta(k)k^{7/5}$ ({\em BO} scaling) and $E_\theta(k)k^{5/3}$ ({\em KO} scaling).  The figures indicate that  for $\mathrm{Ri}=0.01$, {\em BO} scaling fits  with the numerical data better than {\em KO} scaling.
 
\begin{figure}[htbp]
\begin{center}
\includegraphics[scale = 1.0]{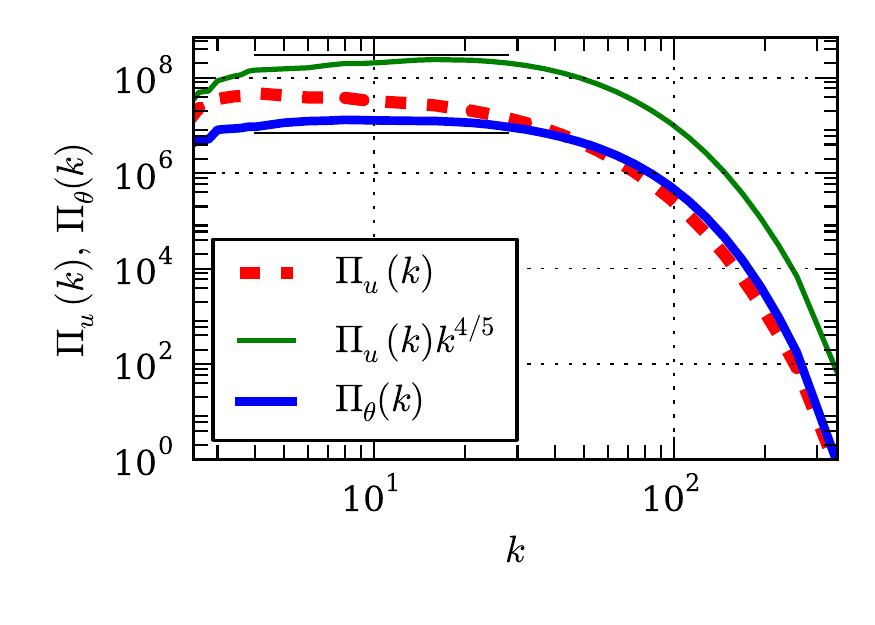}
\end{center}
\caption{For $\mathrm{Pr}=1$ and $\mathrm{Ri} = 0.01$, plots of KE flux $\Pi_u(k)$, normalized KE flux $\Pi_u(k)k^{4/5}$, and entropy flux $\Pi_{\theta}(k)$.  $\Pi_{\theta}(k)$ is multiplied by $10^5$ to fit in the same plot.}
\label{fig:flux}
\end{figure}

\begin{figure}[htbp]
\begin{center}
\includegraphics[scale = 1.0]{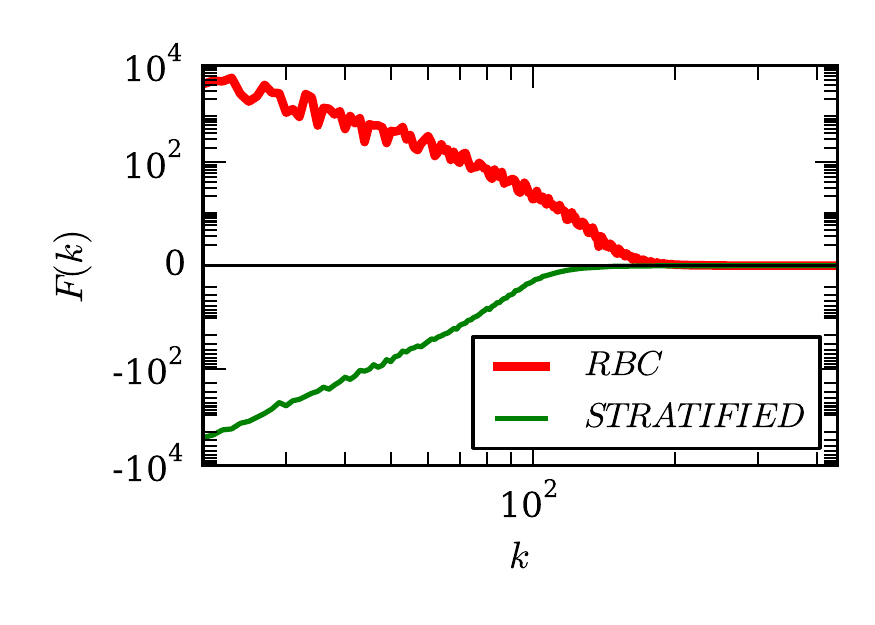}
\end{center}
\caption{Plots of $F(k)$ stably stratified flow ($\mathrm{Pr} = 1$, $\mathrm{Ri} = 0.01$ on $1024^3$ grid), and for  RBC ($\mathrm{Pr} = 1$, $\mathrm{Ra} = 5 \times 10^6$ on $256^3$ grid).  $F(k) < 0$ for stratified flows, but $F(k) > 0$ for RBC.   $F(k)$ for RBC is multiplied by $10^{4}$ to fit in the same plot.}
\label{fig:f_k}
\end{figure}

We cross check our spectrum results with those on KE and entropy fluxes, which are plotted in Fig.~\ref{fig:flux}.   Clearly, the KE flux $\Pi_u(k)$ is positive, and it decreases with $k$.  However  $\Pi_u(k) k^{4/5}$ is almost flat, thus  $\Pi_u(k) \propto k^{-4/5}$, consistent with the {\em BO} predictions [Eq.~(\ref{eq:pi})].    This is consistent with the observed negative $F(k) =  \Re \langle u_z(k) \theta^*(k)\rangle$ for this case (see Fig.~\ref{fig:f_k} and \cite{suppl}).  We also observe that $\Pi_\theta$ is a constant in the inertial range [Eq.~(\ref{eq:pi_theta})].  
 These results show that the {\em BO} scaling is valid for stably stratified flows for  $\rm Ri = O(1)$.   We also compute the Bolgiano wavenumber $k_B$~\cite{Bolgiano:JGR1959} using the numerical data of Eq.~(\ref{eq:pi}), and find that $k_B \approx 8.5$.  Our plots on spectra and fluxes show that  $k_B \approx 8.5$ is only 3 to 4 times smaller than $k_d$, wavenumber where the dissipation range starts.  Therefore a clear-cut crossover from $k^{-11/5}$ to $k^{-5/3}$ is not observed in our simulations.   We are in the process of performing simulations on even higher resolution to probe the crossover region.

\begin{figure}[htbp]
\begin{center}
\includegraphics[scale = 1.0]{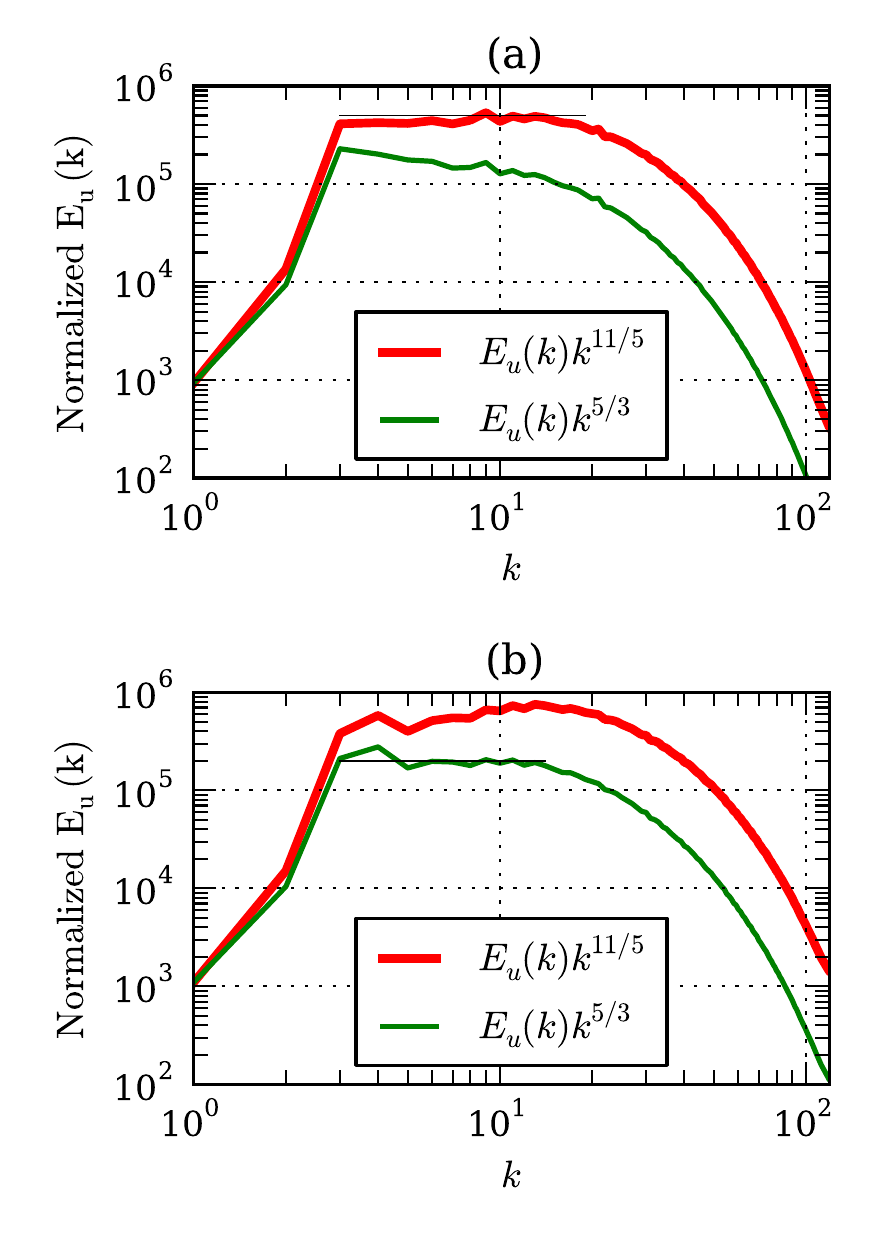}
\end{center}
\caption{Plots of normalized KE spectra for Bolgiano-Obukhov ({\em BO}) scaling and Kolmogorov-Obukhov ({\em KO}) scaling for: (a) $\mathrm{Ri} = 0.5$ and (b) $\mathrm{Ri} = 4 \times 10^{-7}$ }
\label{fig:ke_spec}
\end{figure}

We also performed $512^3$ grid simulations for ${\rm Ri}= 0.5 $ and $4 \times 10^{-7}$ with $\mathrm{Pr}=1$.  The normalized KE spectra for these two cases are exhibited in Figs.~\ref{fig:ke_spec}{\color{blue}(a)} and ~\ref{fig:ke_spec}{\color{blue}(b)} respectively.  Our results show that {\em BO} scaling is valid for ${\rm Ri} = 0.5$, but {\em KO} scaling (with a constant $\Pi_u(k)$) is valid for ${\rm Ri } = 4 \times 10^{-7} $, which is as expected since buoyancy is significant only for moderate and large  ${\rm Ri}$'s.   The energy supply rate by buoyancy, $F(k)$, is significant for ${\rm Ri} = 0.5$, but insignificant for ${\rm Ri } = 4 \times 10^{-7} $, consistent with the above observations~\cite{suppl}.
 
 To contrast the energy supply rate by buoyancy in stratified flows with that in Rayleigh B\'{e}nard convection (RBC), we numerically solve the nondimensionalized RBC equations  for $\mathrm{Pr} = 1.0$ and Rayleigh number $\mathrm{Ra} = 5 \times 10^6$ on $256^3$ grid~\cite{Mishra:PRE2010}. For this run, we plot  $F(k)$ in Fig.~\ref{fig:f_k} that demonstrates that $F(k) > 0$, consistent with our aforementioned arguments, but differs from those of L'vov and Falkovich~\cite{Lvov:PD1992}.  The ongoing work on the flux and spectrum for RBC will be reported in a future work.   Thus, the behaviour of $F(k)$ and $\Pi_u(k)$ for the stably stratified flow and RBC are quite different.
  
\begin{figure}[htbp]
\begin{center}
\includegraphics[scale = 1.0]{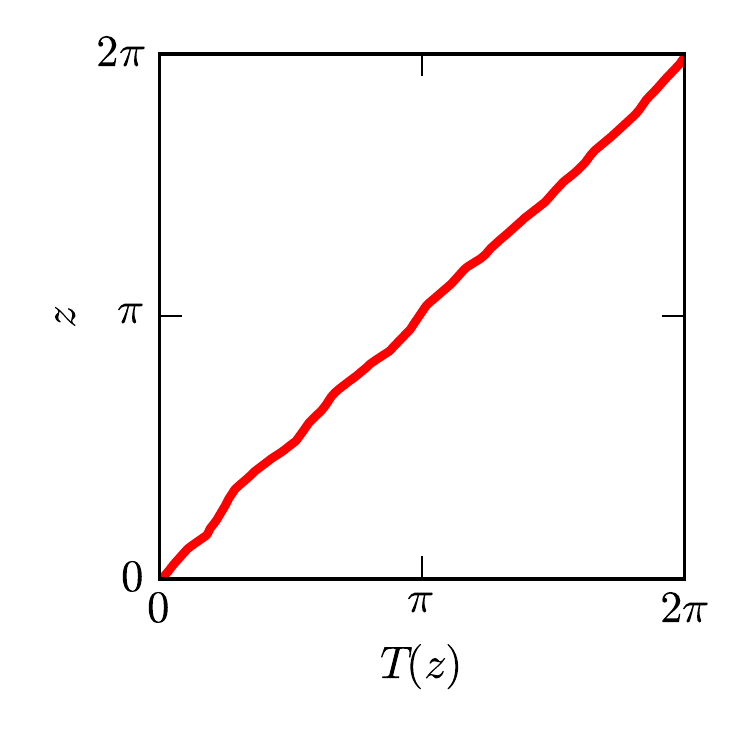}
\end{center}
\caption{The vertical variation of horizontally averaged mean temperature $\bar{T}(z) = \langle T(x,y,z) \rangle_{xy}$ for $\mathrm{Pr}=1$ and $\mathrm{Ri}=0.01$ run.}
\label{fig:temp_pro}
\end{figure}

We employ periodic boundary condition for the stably stratified flows in the vertical direction, thus eliminating the effects of  boundary walls.  In Fig.~\ref{fig:temp_pro} we plot the plane-averaged mean temperature profile $\bar{T}(z) = \langle T(x,y,z) \rangle_{xy}$.   Since $\bar{T}(z) $ is linear,  a constant temperature gradient $d\bar{T}/dz$ (hence buoyancy) acts in the whole box.  Therefore,  {\em BO} scaling is expected everywhere.   It is important to contrast the above profile with that for Rayleigh-B\'{e}nard convection in which most of the temperature drop takes place in a narrow thermal boundary layer~\cite{Moore:JFM1973,Verzicco:JFM2003}, while the bulk flow has $d\bar{T}/dz \approx 0$.  Thus we expect {\em BO} scaling in the boundary layers, and {\em KO} scaling in the bulk, as reported by Calzavarini et al.~\cite{Calzavarini:PRE2002}.

In summary, our numerical simulations demonstrate an existence of {\em BO} scaling in stably stratified flows.  A major novelty in our approach is the quantitative analysis of the KE and entropy fluxes, as well as the energy supply rate by buoyancy ($F(k)$).   We show that $F(k) < 0$ for stably stratified flows, but $F(k) > 0$ for Rayleigh B\'{e}nard convection.  Consequently, for stably stratified flows with moderate Richardson numbers,  the energy flux $\Pi_u(k) \propto k^{-4/5}$ and $E_u(k) \sim k^{-11/5}$, as proposed in the {\em BO} phenomenology.  However,  $F(k)$ is somewhat insignificant for small Richardson number, and we observe {\em KO} scaling.  For RBC flows, $\Pi_u(k)$ is a non-decreasing function of $k$ (since $F(k) > 0$), and the energy  spectrum of KE cannot be steeper than $k^{-5/3}$ in the bulk.  Thus, energy flux and energy supply rate due to buoyancy provide  valuable insights into the physics of stably stratified flows and RBC.

Our numerical simulations were performed at Centre for Development of Advanced Computing (CDAC) and IBM Blue Gene P ``Shaheen" at KAUST supercomputing laboratory, Saudi Arabia. This work was supported by a research grant SERB/F/3279/2013-14 from Science and Engineering Research Board, India. We thank Ambrish Pandey, Anindya Chatterjee, Pankaj Mishra, and Mani Chandra for valuable suggestions.


\pagebreak
\section{Supplemental Material}

In Fourier space, the equation for the kinetic energy  (KE) can be derived from Eq.~(1) of the main paper as~\cite{Lesieur:book,Lvov:PRL1991,Verma:EPL2012}
\begin{equation}
\frac{\partial E_u(k)}{\partial t} = T_u(k) + F(k) - D(k)
\label{eq:E}
\end{equation}
where $T_u(k)$ is energy transfer rate to a wavenumber shell of radius $k$, and it is related to the energy flux of KE as
\begin{equation}
\Pi_u(k) = - \int_0^k T_u(k)\,dk.
\label{eq:Pi}
\end{equation}
The energy supply rate $F(k)$ of Eq.~(\ref{eq:E}) is given by
 \begin{equation}
F(k) = \mathrm{Gr Pr^2} \sum_{|{\mathbf k}| = k} \Re\langle u_z({\mathbf k}) \theta^*({\mathbf k}) \rangle + \Re\langle {\mathbf u}({\mathbf k}) \cdot {\mathbf f}^*({\mathbf k}) \rangle,
\label{eq:F}
\end{equation}
 where the first term is due to buoyancy, while the second term is due to the external random forcing, which is active only for $2 \le k \le 4$.   The viscous dissipation is given by
 \begin{equation}
D(k) = \sum_{|{\mathbf k}| = k} 2 \mathrm{Pr} k^2 E_u(k)
\label{eq:Pi}
\end{equation}
 From the above equation, we deduce that 
\begin{equation}
\frac{d}{dk}\Pi(k) = -T(k) = -\frac{\partial E}{\partial t} + F(k) - D(k).
\label{eq:dPi}
\end{equation}
Under a steady state ($\partial E/\partial t =0$), we obtain
\begin{equation}
\frac{d}{dk}\Pi(k) = F(k) - D(k).
\label{eq:dPi}
\end{equation}
In our simulations of stably stratified flows, the external force  ${\mathbf f}^*({\mathbf k})$ is active for the band $2 \le k \le 4$.  Therefore, we focus on wavenumbers $k > 4$ where only buoyancy force is active.

\begin{figure}[htbp]
\begin{center}
\includegraphics[scale = 0.22]{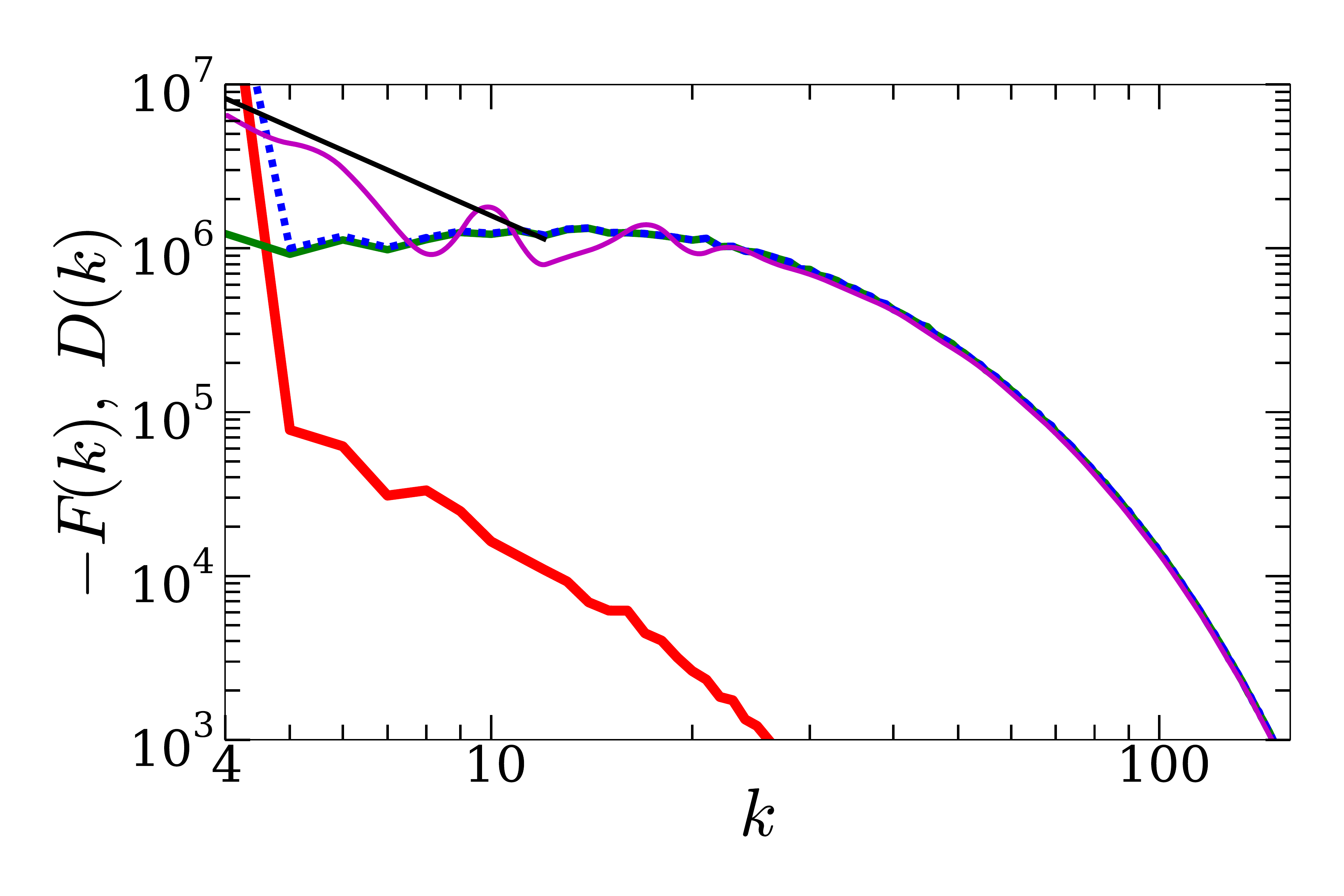}
\end{center}
\caption{For $\mathrm{Pr}=1$ and $\mathrm{Ri}=0.01$, plots of $-F(k)$ (thick red curve), $D(k)$ (thick green curve), $[-F(k)+D(k)]$ (dotted blue curve) , and $-\frac{d}{dk}\Pi_u(k)$ (thin magenta curve). $k^{-9/5}$ scaling is shown using black line.}
\label{fig:uz_theta_ri_0.01}
\end{figure}

\begin{figure}[htbp]
\begin{center}
\includegraphics[scale = 0.22]{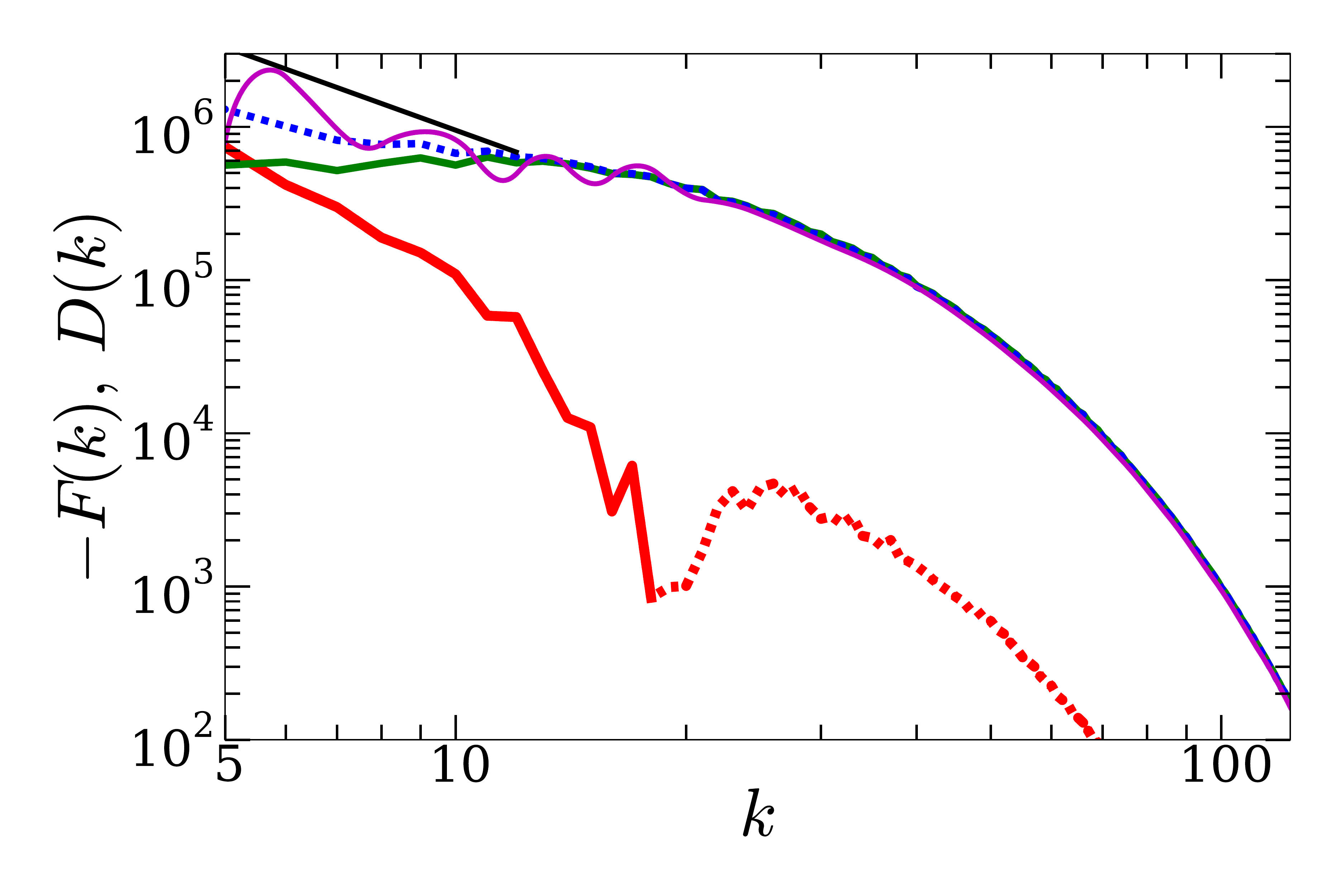}
\end{center}
\caption{For $\mathrm{Pr}=1$ and $\mathrm{Ri}=0.5$, $F(k)$ has  both signs.  We exhibit negative $F(k)$ (lower $k$) with thick red curve, and   positive $F(k)$ (higher $k$)  using thick dotted red curve.   We also exhibit $D(k)$ (thick green curve), $[-F(k)+D(k)]$ (thick dotted blue curve), and $-\frac{d}{dk}\Pi_u(k)$ (thin magenta curve).  $k^{-9/5}$ scaling is shown using black line.}
\label{fig:uz_theta_ri_0.5}
\end{figure}

\begin{figure}[htbp]
\begin{center}
\includegraphics[scale = 0.22]{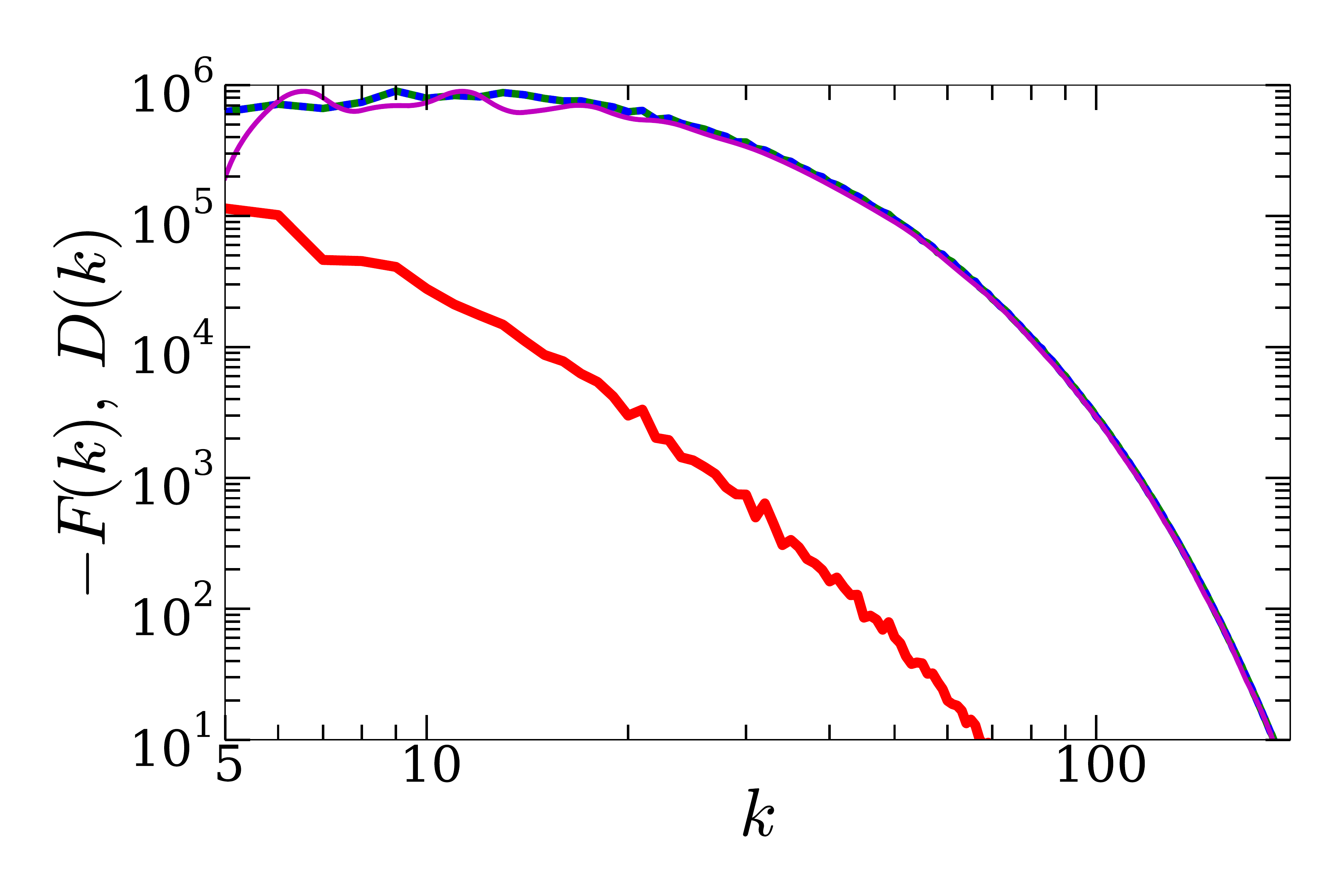}
\end{center}
\caption{For $\mathrm{Pr}=1$ and $\mathrm{Ri}=4 \times 10^{-7}$, plots of $-F(k)$ (thick red curve), $D(k)$ (thick green curve), $[-F(k)+D(k)]$ (dotted blue curve) , and $-\frac{d}{dk}\Pi_u(k)$ (thin magenta curve). $F(k)$ is multiplied by $10^5$ to fit in the same plot.}
\label{fig:uz_theta_ri_4e-7}
\end{figure}

We compute $F(k)$ and $D(k)$ using the numerical data computed for stably stratified flows with $\mathrm{Pr}=1$, and  $\mathrm{Ri}=0.01, 0.5,4\times 10^{-7}$.  These quantities are shown in Figs.~\ref{fig:uz_theta_ri_0.01}, \ref{fig:uz_theta_ri_0.5}, and \ref{fig:uz_theta_ri_4e-7} respectively.   In the inertial range, $F(k)$ is negative for all the three cases, consistent with the predictions of Bolgiano-Obukhov ({\em BO}) phenomenology.   We observe that  for $\mathrm{Ri}=0.01$ and  $0.5$, $F(k)$ is significant for small wavenumbers. However for   $\mathrm{Ri}=4\times 10^{-7}$, buoyancy is weak, and $|F(k)| \ll D(k)$.  These results are consistent with the flux and spectra results presented in the main paper.

We also find that $|F(k)|$ drops sharply with $k$ for $\mathrm{Ri}=0.01, 0.5$.  Since $\Pi(k) \sim k^{-4/5}$, we observe $F(k) \sim d\Pi/dk \sim -k^{-9/5}$ for a narrow band in the small wavenumber regime.   In contrast, for $\mathrm{Ri}=4\times 10^{-7}$, $F(k)$ is much smaller than the corresponding $F(k)$ for $\mathrm{Ri}=0.01, 0.5$, consistent with $d\Pi(k)/dk \approx -D(k)$.

\begin{figure}[htbp]
\begin{center}
\includegraphics[scale = 0.22]{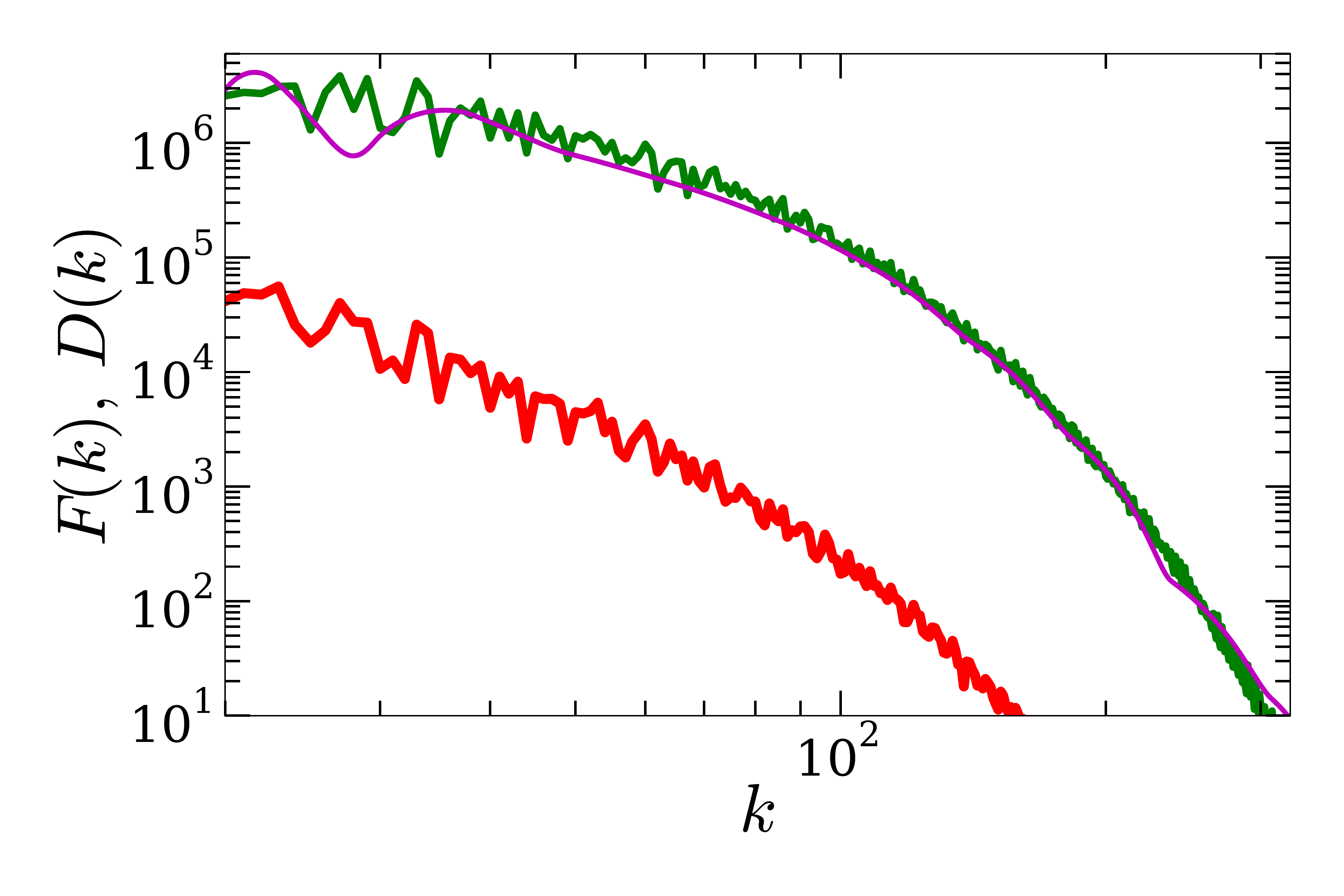}
\end{center}
\caption{For $\mathrm{Pr}=1$ and $\mathrm{Ra}=5 \times 10^6$, plots of $F(k)$ (thick red curve), $D(k)$ (thick green curve), and $-\frac{d}{dk}\Pi_u(k)$ (thin magenta curve) on $256^3$ grid. $F(k)$ is multiplied by $10^5$ to fit in the same plot.}
\label{fig:uz_theta_c}
\end{figure}

We contrast the above results with those for  Rayleigh B\'{e}nard convection (RBC).  We numerically solve the nondimensionalized RBC equations under the Boussinesq approximation  for $\mathrm{Pr} = 1.0$ and Rayleigh number $\mathrm{Ra} = 5 \times 10^6$ on $256^3$ grid~\cite{Mishra:PRE2010}. For this run, we plot  $F(k)$ and $D(k)$ in Fig.~\ref{fig:uz_theta_c} that demonstrates that $F(k) > 0$, consistent with our arguments.  Note however that for this case, $F(k) \ll D(k)$ and $d\Pi_u(k)/dk = -D(k)< 0$, or $\Pi_u(k)$ is decreasing with $k$ due to dominance of $D(k)$ over $F(k)$.  We need to perform very large-resolution simulation for much higher $\mathrm{Ra}$ that would provide significant inertial range where $\Pi_u(k)$ could be a non-decreasing function of $k$.   
  
\end{document}